\documentclass[12pt]{article}

\usepackage{graphicx}
\usepackage{amsmath, amssymb}

\def\calp{{\mathcal P}}
\def\fq{\mathbb{F}_{q}}
\def\fd{\mathbb{F}_{2}}
\def\nm{\mathbb{N}}

\begin{document}

\title{\bf Finite geometries and diffractive orbits in isospectral billiards}
\author{Olivier Giraud\\ 
Laboratoire de Physique Th\'eorique,\\
 UMR 5152 du CNRS, \\
Universit\'e Paul Sabatier,\\
 31062 Toulouse Cedex 4,\\
France\\}
\maketitle

\begin{abstract}
Several examples of pairs of isospectral planar domains have been produced
in the two-dimensional Euclidean space by various methods. 
We show that all these examples rely on the symmetry between 
points and blocks in finite projective spaces; from the properties 
of these spaces, one can derive a relation between Green functions 
as well as a relation between diffractive orbits in isospectral billiards.
\end{abstract}

\begin{figure}[ht]
\begin{center}
\includegraphics[width=0.68\linewidth]{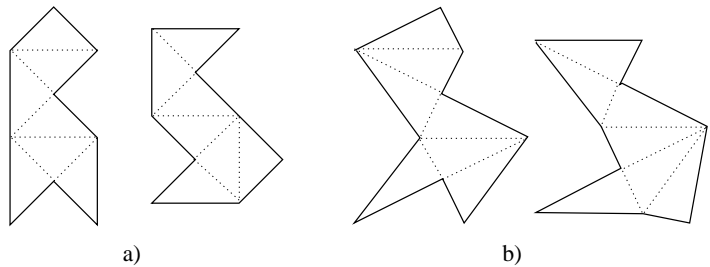}
\end{center}
\caption{a) Two isospectral billiards with a triangular base shape.
b) A pair of isospectral billiards constructed from the same unfolding rules 
as a).}
\label{billards}
\end{figure}

It is almost forty years now that Mark  Ka{\v c} addressed his 
famous question ``Can one hear the shape of a drum'' \cite{Kac66}. 
The problem was to know whether there exists
 isospectral domains, that is non-isometric bounded regions of 
space for which the sets $\{E_n, n\in\nm\}$ of solutions of the 
stationary Schr\"odinger equation $(\Delta+E)\Psi=0$, with some 
specified boundary conditions, would coincide.
Instances of isospectral pairs have finally been found for Riemannian manifolds
 \cite{Be89, Br88}, and more recently for two-dimensional Euclidean 
connected compact domains (we will call such planar domains ''billiards'') 
\cite{GorWebWol92}.
The proof of isospectrality was based on Sunada's theorem, which allows to 
construct isospectral pairs from groups related by a so-called 
''almost conjugate'' property \cite{Sun85}. 
The two billiards given in \cite{GorWebWol92} to illustrate the existence of
isospectral billiards are represented in \ref{billards}a: they are made 
of 7 copies of a base tile (here a right isosceles triangle) unfolded with
respect to its edges in two different ways. 
It turns out that in this example, isospectrality can be 
proved directly by giving an explicit linear map between the two 
domains \cite{Cha95}. Eigenfunctions in one billiard can be expressed as a
linear superposition of eigenfunctions of the other billiard, in such
a way that boundary conditions are fulfilled; isospectrality is
ensured by linearity of Schr\"odinger equation. This 
''transplantation'' method allowed to generalize the example of 
\cite{GorWebWol92} : any triangle can replace 
the base triangle of \ref{billards}a, what matters is the way 
the initial base tile is unfolded (see \ref{billards}b).
Further examples of isospectral billiards were obtained by applying 
Sunada's theorem to reflexion groups in the 
hyperbolic upper-half plane \cite{BusConDoySem94}: examples of
billiards made of 7, 13, 15 or 21 triangular tiles were given.
Later on, it was shown that the transplantation 
method produces exactly the same examples and provides 
a proof for iso-length-spectrality of these billiards
(i.e. the set of lengths of periodic orbits is identical) \cite{OkaShu01}.
A similar transplantation
technique, together with results on Green functions of polygonal billiards,
was applied to show that there is no such equality of the lengths for
diffractive orbits (that is, the lines joining two vertices); a
more involved relation was found \cite{Gir04}.\\

The purpose of this letter is to show that all known examples of
isospectral pairs, given by \cite{BusConDoySem94} or \cite{OkaShu01}, 
can be obtained by considering finite projective spaces
(FPS): the transplantation map between two isospectral billiards 
can be taken to be the incidence matrix of a FPS. This construction 
from FPSs indicates that the deep origin of isospectrality in 
Euclidean spaces is point-block duality. It also leads to a
relation between Green functions, and to a correspondence 
between diffractive orbits in isospectral billiards.

\subsection*{Isospectral billiards.} 
\begin{figure}[ht]
\begin{center}
\includegraphics[width=0.66\linewidth]{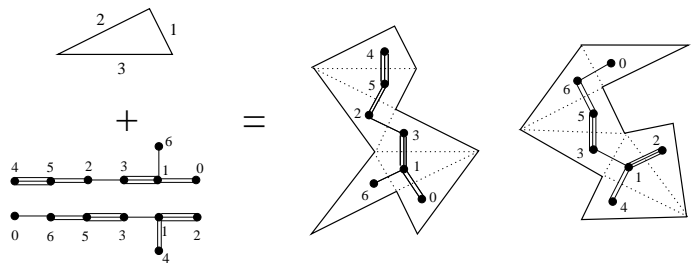}
\end{center}
\caption{The graphs corresponding to a pair of isospectral billiards: 
if we label the sides of the triangle by $\mu=1,2,3$, the unfolding rule
by symmetry with respect to side $\mu$ can be represented by edges made 
of $\mu$ braids in the graph. From a given pair of graphs, one can construct
infinitely many pairs of isospectral billiards by applying the unfolding
rules to any triangle. Note that a different labeling of the tiles would just
induce a permutation of the labelings of points and blocks in the Fano plane.}
\label{graphs}
\end{figure}

All known isospectral billiards can be obtained by unfolding 
triangle-shaped tiles \cite{BusConDoySem94, OkaShu01}.
The way the tiles are unfolded can be specified by three 
permutation matrices $M^{(\mu)}$, $1\leq \mu\leq 3$, associated to the 
three sides of the triangle: $M^{(\mu)}_{ij}=1$ if tiles $i$ and $j$
are glued by their side $\mu$ (and $M^{(\mu)}_{ii}=1$ if the side $\mu$
of tile $i$ is the boundary of the billiard), 
and 0 otherwise \cite{OkaShu01, Tha04, Gir04}.
Following \cite{OkaShu01}, one can sum up the action of the $M^{(\mu)}$
in a graph with coloured edges: each copy of the base 
tile is associated to a vertex, and vertices $i$ and $j$, $i\neq j$,
are linked by an edge of colour $\mu$ if and only if $M^{(\mu)}_{ij}=1$ 
(see \ref{graphs}).
In the same way, in the second member of the pair, the tiles are 
unfolded according to
permutation matrices $N^{(\mu)}$, $1\leq \mu\leq 3$. Two billiards are
said to be transplantable if there exists an invertible matrix $T$
(the {\it transplantation matrix})
such that $\forall\mu\ \ T M^{(\mu)}= N^{(\mu)} T$. One can show that
transplantability implies isospectrality (if the matrix $T$ is not 
merely a permutation matrix, in which case the two domains would just
have the same shape). The underlying 
idea is that if $\psi^{(1)}$ is
an eigenfunction of the first billiard and $\psi^{(1)}_i$ its restriction
to triangle $i$, then one can build an eigenfunction  $\psi^{(2)}$ 
of the second billiard by taking $\psi^{(2)}_i=\sum_j T_{ij}\psi^{(1)}_j$. 
Obviously $\psi^{(2)}$
verifies Schr\"odinger equation; it can be checked from the
commutation relations that the function is smooth at all edges 
of the triangles, and that boundary conditions at the boundary of 
the billiard are fulfilled \cite{OkaShu01}.\\

Suppose we want to construct a pair of isospectral billiards, starting
from any polygonal base shape. Our idea is to start from the 
transplantation matrix, and choose it in such a way that the existence
of commutation relations $T M^{(\mu)}= N^{(\mu)} T$ for some 
permutation matrices $M^{(\mu)}, N^{(\mu)}$ will be known {\it a priori}.
As we will see, this is the case if $T$ is taken to be
the incidence matrix of a FPS; the matrices $M^{(\mu)}$ and
$N^{(\mu)}$ are then permutations on the points and the hyperplanes
of the FPS.

\subsection*{Finite projective spaces.} 
\begin{figure}[ht]
\begin{center}
\includegraphics[width=0.58\linewidth]{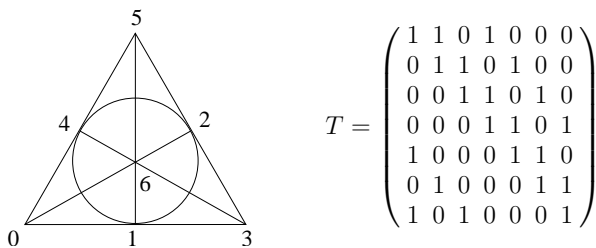}
\caption{The Fano plane $PG(2,2)$ and its corresponding incidence
matrix $T$. The Fano plane has $(q^3-1)/(q-1)=7$ points 
and 7 lines; each line contains $q+1=3$ points and each point belongs 
to 3 lines. Any pair of points belongs to one and only one line.}
\label{fano}
\end{center}
\end{figure}

For $n\geq 2$ and $q=p^h$  a power of a prime number, consider 
the $(n+1)$-dimensional vector space $\fq^{n+1}$, where
$\fq$ is the finite field of order $q$.
The {\it finite projective space} $PG(n,q)$ of {\it dimension} $n$
and {\it order} $q$ is the set of subspaces of $\fq^{n+1}$: 
the points of $PG(n,q)$ are the 1-dimensional subspaces of $\fq^{n+1}$, 
the lines of $PG(n,q)$ are the 2-dimensional subspaces of $\fq^{n+1}$,
and more generally $(d+1)$-dimensional subspaces of $\fq^{n+1}$ are 
called {\it $d$-spaces} of $PG(n,q)$; the $(n-1)$-spaces of $PG(n,q)$ 
are called hyperplanes or {\it blocks}. 
A $d$-space of $PG(n,q)$ contains $(q^{d+1}-1)/(q-1)$ points. In particular,
$PG(n,q)$ has $(q^{n+1}-1)/(q-1)$ points. It also has 
$(q^{n+1}-1)/(q-1)$ blocks \cite{Hir79}.
As an example, \ref{fano} shows the finite projective plane (FPP) 
of order $q=2$, or Fano plane, $PG(2,2)$. 

A $(N,k, \lambda)-${\it symmetric balanced incomplete block design} 
(SBIBD) is a set of
$N$ points, belonging to $N$ subsets (or {\it blocks});
each block contains $k$ points, in such a way that any two points belong 
to exactly $\lambda$ blocks, and each point is contained in 
$k$ different blocks \cite{DinSti92}. One can show that
$PG(n,q)$ is a $(N,k, \lambda)$-SBIBD with $N=(q^{n+1}-1)/(q-1)$,
$k=(q^{n}-1)/(q-1)$ and $\lambda=(q^{n-1}-1)/(q-1))$. 
For example, the Fano plane is a $(7,3,1)-$SBIBD.
The points and the blocks can be labeled from $0$ to $N-1$.
For any $(N,k, \lambda)-$SBIBD one can define an $N\times N$
{\it incidence matrix} 
$T$ describing to which block each point belongs. The entries $T_{ij}$ 
of the matrix are $T_{ij}=1$ if point $j$ belongs to line $i$, $0$ 
otherwise. The matrix $T$ verifies the relation
$T T^{t}=\lambda J+(N-k)\lambda/(k-1)I$, 
where $J$ is the $N\times N$ matrix with all entries equal to 1 
and $I$ the $N\times N$ identity matrix \cite{DinSti92}. 
In particular, the incidence matrix of $PG(n,q)$ verifies
\begin{equation}
\label{tt}
T T^{t}=\lambda J+(k-\lambda)I
\end{equation}
with $k$ and $\lambda$ as given above. For example, the incidence matrix 
of the Fano plane given in 
\ref{fano} corresponds to a labeling of the lines such that line 0
contains points 0,1,3, and line 1 contains points 1,2,4, etc.

A {\it collineation} of a FPS is a 
bijection that preserves incidence, that
is a permutation of the points that takes $d$-spaces to 
$d$-spaces (in particular, it takes blocks to blocks).
Any permutation $\sigma$ on the points
can be written as a $N\times N$ {\it permutation matrix} $M$ defined by 
$M_{i\sigma(i)}=1$ and the other entries equal to zero. If $M$ is a 
permutation matrix associated to a collineation, then there exists a 
permutation matrix $N$ such that
\begin{equation}
\label{commutation}
TM=NT.
\end{equation}
In words, (\ref{commutation}) means that permuting the 
columns of $T$ (i.e. the blocks of the
space) under $M$ is equivalent to permuting the rows of $T$ 
(i.e. the points of the space) under $N$.
The commutation relation (\ref{commutation}) is a 
related to an important feature of projective geometry, the so-called
''principle of duality'' \cite{Hir79}. 
This principle states that for any theorem which is 
true in a FPS, the dual theorem obtained by exchanging 
the words ''point'' and ''block'' is also true. As we will see now, 
this symmetry between points and blocks in FPSs is the central reason 
which accounts for known pairs of isospectral billiards.\\

Let us consider a FPS $\calp$ with incidence matrix $T$.
To each block in $\calp$ we associate a tile in the first billiard, 
and to each point in $\calp$ we associate a tile in the second billiard.
If we choose permutations $M^{(\mu)}$ in the collineation 
group of $\calp$, then the commutation relation (\ref{commutation})
will ensure that there exist permutations $N^{(\mu)}$ verifying
$TM^{(\mu)}=N^{(\mu)}T$. These commutation relations imply 
transplantability, and thus isospectrality, of the billiards 
constructed from the graphs corresponding to $M^{(\mu)}$ and $N^{(\mu)}$.

If the base tile has $r$ sides, we need to choose $r$ 
elements $M^{(\mu)}$, $1\leq\mu\leq r$, in the 
collineation group of $\calp$. This choice is 
constrained by several factors. Since
$M^{(\mu)}$ represents the reflexion of a tile with respect to one of
its sides, it has to be of order 2 (i.e. an involution). 
In order that the billiards be connected, no point should be left out by 
the matrices $M^{(\mu)}$ (in other words, the graph associated to the
matrices $M^{(\mu)}$ should be connected). 
Finally, if we want the base tile to be of any shape, there should 
be no loop in the graph. We now need to characterize 
collineations of order 2.

\subsection*{Collineations of finite projective spaces.} 
Let $q=p^h$ be a power of a prime number.
Each point $P$ of $PG(n,q)$ is a 1-dimensional subspace of $\fq^{n+1}$, 
spanned by some vector $v$. We write $P=P(v)$.

An {\it automorphism} is a bijection of the points $P(v_i)$
of $PG(n,q)$ obtained 
by the action of an automorphism of $\fq$ on the coordinates 
of  the $v_i$. If $q=p^h$, the automorphisms of $\fq$ are
$t\mapsto t^{p^i}$, $0\leq i \leq h-1$.

A {\it projectivity} is a bijection of the points $P(v_i)$ 
of $PG(n,q)$ obtained by the action of a linear map $L$ on the $v_i$.
There are $q-1$ matrices $t L$, with $t\in\fq\setminus\{0\}$ and 
$L\in GL_{n+1}(\fq)$ (the group of $(n+1)\times (n+1)$ invertible
matrices with coefficients in $\fq$), yielding the same projectivity
$P(v_i)\mapsto P(L v_i)$.

The {\it Fundamental theorem of projective geometry } states
that any collineation of $PG(n,q)$ can be written as
the composition of a projectivity by an automorphism \cite{Hir79}.
The converse is obviously true since projectivities and automorphisms 
are collineations. Therefore the set of all collineations is obtained
by taking the composition of all the non-singular $(n+1)\times(n+1)$
matrices with coefficients in $\fq$ by all the $h$ automorphisms of $\fq$.

The collineation group of $PG(n, q)$ has 
$[h\prod_{k=0}^{n}(q^{n+1}-q^k)]/(q-1)$ elements, among which we only
want to consider elements of order 2. In the case of FPPs ($n=2$), 
there are various known properties characterizing
collineations of order 2. 
A {\it central collineation}, or {\it perspectivity}, is a collineation fixing
each line through a point $C$ (called the centre). By ''fixing'' we
mean that the line is invariant but the points can be permuted within the
line. One can show that the fixed points of a non-identical perspectivity
are the centre itself
and all points on a line (called the axis), while the fixed lines are
the axis and all lines through the centre. 
If the centre lies off the axis a perspectivity is called a 
{\it homology} (and has $q+2$ fixed points), whereas if the centre 
lies on the axis it is called an {\it elation} (and has $q+1$ fixed points).
Perspectivities in dimension $n=2$ have following properties \cite{Bon04}:

{\scshape Proposition 1.} A perspectivity of order two 
of a FPP of order $q$ is an elation or a 
homology according to whether $q$ is even or odd.

{\scshape Proposition 2.} A collineation of order two of a FPP
 of order $q$ is a perspectivity if $q$ 
is not a square; it is a collineation fixing all points and lines 
in a subplane if $q$ is a square (a subplane is a subset of points 
having all the properties of a FPP).

When the order of the FPP is a square, there is 
the following result \cite{Hir79}:

{\scshape Proposition 3.} $PG(2, q^2)$ can be partitioned into $q^2-q+1$
subplanes $PG(2, q)$.

\subsection*{Generating isospectral pairs.} 
Let us assume we are looking for a pair of isospectral billiards with
$N=(q^3-1)/(q-1)$ copies of a base tile having the 
shape of a $r$-gon, $r\geq 3$. 
We need to find $r$ collineations of order 2 such that the associated graph is
connected and without loop. Such a graph connects $N$ vertices and thus
 requires $N-1$ edges. From propositions 1-3, we can deduce the 
number $s$ of fixed points of a collineation of order 2 for any FPP. 
Since a collineation is a permutation, it has a cycle decomposition as
a product of transpositions. For permutations of order 2 with $s$ fixed 
points, there are $(N-s)/2$ independent transpositions in this decomposition.
Each transposition is represented by an edge in the graph. As a consequence,
 $q$, $r$ and $s$ have to fulfill the following condition: 
$r(q^2+q+1-s)/2=q^2+q$. Let us examine the various cases.

{\it If $q$ is even and not a square},
propositions 1 and 2 imply that 
any collineation of order 2 is an elation and therefore has $q+1$
fixed points. Therefore, $q$ and $r$ are constrained by the relation
$r q^2/2=q^2+q$.
The only integer solution with $r\geq 3$ and $q\geq 2$ is $(r=3, q=2)$. 
These isospectral billiards  correspond to $PG(2,2)$ and will be made
of $N=7$ copies of a base triangle.

{\it If $q$ is odd and not a square},
propositions 1 and 2 imply that 
any collineation of order 2 is a homology and therefore has $q+2$
fixed points. Therefore, $q$ and $r$ are constrained by the relation
$r (q^2-1)/2=q^2+q$.
The only integer solution with $r\geq 3$ and $q\geq 2$ is $(r=3, q=3)$. 
These isospectral billiards correspond to $PG(2,3)$ and will be made
of $N=13$ copies of a base triangle.

{\it If $q=p^2$ is a square}, 
propositions 2 and 3 imply that 
any collineation of order 2 fixes all points in a subplane $PG(2,p)$
and therefore has $p^2+p+1$ fixed points. Therefore, $p$ and $r$ are 
constrained by the relation $r (p^4-p)/2=p^4+p^2$.
There is no integer solution with $r\geq 3$ and $q\geq 2$. However,
one can look for isospectral billiards with loops: this will require
the base tile to have a shape such that the loop  does not make the
copies of the tiles come on top of each other when unfolded. 
If we tolerate one loop
in the graph describing the isospectral billiards, then there are $N$ 
edges in the graph instead of $N-1$ and the equation for $p$ and $r$
becomes $r (p^4-p)/2=p^4+p^2+1$, which has the only integer solution
$(r=3, p=2)$. These isospectral billiards correspond to $PG(2,4)$
and will be made of $N=21$ copies of a base triangle.

We can now generate all possible pairs of isospectral billiards 
whose transplantation matrix is the incidence matrix of $PG(2,q)$, 
with  $r$ and $q$ restricted by the previous analysis. All pairs 
must have a triangular base shape ($r=3$).
$PG(2, 2)$ provides 3 pairs (made of 7 tiles), 
$PG(2, 3)$ provides 9 pairs (made of 13 tiles), 
$PG(2, 4)$ provides 1 pair (made of 21 tiles).
It turns out that the pairs obtained here are exactly those obtained by 
\cite{BusConDoySem94} and \cite{OkaShu01} by other methods.
Let us now consider spaces $PG(n,q)$ of higher dimensions. The smallest
one is $PG(3,2)$, which contains 15 points.
Since the base field for $PG(3,2)$ is $\fd$, the Fundamental 
Theorem of projective geometry states that the collineation group of 
$PG(3,2)$ is essentially the group $GL_4(\fd)$ of 
$4\times 4$ non-singular matrices with coefficients in $\fd$. 
Generating all possible graphs from the 316
elements of order 2 in $GL_4(\fd)$, we obtain four pairs of 
isospectral billiards with 15 triangular tiles, which completes
the list of all pairs found in \cite{BusConDoySem94} and \cite{OkaShu01}.\\

Our method explicitly gives the transplantation matrix $T$ for all these
pairs: each one is the incidence matrix of a FPS. 
The transplantation matrix explicitly provides the mapping between
eigenfunctions of both billiards. The inverse mapping is given by 
$T^{-1}=(1/q^{n-1})(T^{t}-(\lambda/k)J)$.
For all pairs, isospectrality can therefore be explained 
by the symmetry between points and blocks in FPSs.
We do not know if it is possible to find isospectral billiards for 
which isospectrality would not rely on this symmetry.

Our construction furthermore allows to generalize the results we obtained 
in \cite{Gir04}, where a relation between the Green functions of the 
billiards in an isospectral pair was derived. 
A similar relation can be found for all other pairs
obtained by point-block duality. Let $M^{(\mu)}$ and $N^{(\mu)}$
be the matrices describing the gluings of the tiles.
To any path $p$ going from a point to another on the first billiard, 
one can associate the sequence $(\mu_1,...,\mu_n)$, $1\leq\mu_i\leq 3$,
of edges of the triangle hit by the path. The matrix $M=\prod M^{(\mu_i)}$
is such that $M_{ij}=1$ if path $p$ can be drawn from $i$ to $j$.
If $N=T^{-1}MT$, relations (\ref{tt}) and (\ref{commutation}) imply  
$\sum_{k,l}T_{ik}T_{jl}M_{kl}=\lambda+(k-\lambda)N_{ij}$. 
Since Green functions can be written as a sum
over all paths, the relation between the Green functions $G^{(2)}(a,i;b,j)$ 
and $G^{(1)}(a,i';b,j')$ is
$\sum_{i',j'}T_{i,i'}T_{j,j'}G^{(1)}(a,i';b,j')=
(k-\lambda)G^{(2)}(a,i;b,j)+\lambda G^{t}(a;b)$, where $G^{t}$ is the 
Green function of the triangle, and a point $(a,i)$ is specified 
by a tile number $i$ and its position $a$ inside the tile
(see \cite{Gir04} for further detail).  
More precisely, the term $\sum_{k,l}T_{ik}T_{jl}M_{kl}$
 can be interpreted as the number of pairs of
tiles $(i,j)$ in the first billiard such that a path identical to
$p$ can go from $i$ to $j$.  
A given diffractive orbit going from $i$ to $j$ in the second 
billiard corresponds to a matrix $N$ such that $N_{ij}=1$: 
it is therefore constructed from a superposition
of $k=(q^n-1)/(q-1)$ identical diffractive orbits in the first billiard.
(Note that these results correspond to Neumann boundary conditions. It is
easy to obtain similar relations for Dirichlet
boundary conditions by
conjugating all matrices with a diagonal matrix $D$ with entries 
$D_{ii}=\pm 1$ according to whether tile $i$ is like the initial tile 
or like its mirror inverse.)


\end{document}